\documentclass[10pt,conference]{IEEEtran}

\usepackage{cite}
\usepackage{amsmath,amssymb,amsfonts}
\usepackage{algorithmic}
\usepackage{authblk}
\usepackage{graphicx}
\usepackage{textcomp}
\usepackage{xcolor}
\usepackage{tcolorbox}
\usepackage{enumitem}
\tcbuselibrary{skins}
\def\BibTeX{{\rm B\kern-.05em{\sc i\kern-.025em b}\kern-.08em
    T\kern-.1667em\lower.7ex\hbox{E}\kern-.125emX}}
\begin{document}

\bstctlcite{IEEEexample:BSTcontrol}

\title{
Beyond the Classroom: Bridging the Gap Between Academia and Industry with a Hands-on Learning Approach
}

\author[1]{Mingyang Xu}
\author[1]{Ryan Liu}
\author[2]{Mark Stoodley}
\author[1]{Ladan Tahvildari}
\affil[1]{Dept. of Electrical and Computer Engineering, University of Waterloo, Canada}
\affil[2]{IBM Canada}
\affil[ ]{\textit{\{mingyang.xu, ryan.zheng.he.liu, ladan.tahvildari\}@uwaterloo.ca}}
\affil[ ]{\textit{mstoodle@ca.ibm.com}}

\maketitle

\begin{abstract}

Modern software systems require various capabilities to meet architectural and operational demands, such as the ability to scale automatically and recover from sudden failures. Self-adaptive software systems have emerged as a critical focus in software design and operation due to their capacity to autonomously adapt to changing environments. However, educating students on this topic is scarce in academia, and a survey among practitioners identified that the lack of knowledgeable individuals has hindered its adoption in the industry. In this paper, we present our experience teaching a course on self-adaptive software systems that integrates theoretical knowledge and hands-on learning with industry-relevant technologies. To close the gap between academic education and industry practices, we incorporated guest lectures from experts and showcases featuring industry professionals as judges, improving technical and communication skills for our students. Feedback based on surveys from 21 students indicates significant improvements in their understanding of self-adaptive systems. The empirical analysis of the developed course demonstrates the effectiveness of the proposed course syllabus and teaching methodology. In addition, we provide a summary of the educational challenges of running this unique course, including balancing theory and practice, addressing the diverse backgrounds and motivations of students, and integrating the industry-relevant technologies. We believe these insights can provide valuable guidance for educating students in other emerging topics within software engineering.

\end{abstract}


\section{Introduction}
Over the past decades, software systems have become increasingly complex, driving significant interest in developing autonomous frameworks capable of meeting evolving operational requirements. Traditionally, such tasks have been managed by dedicated operational teams within organizations~\cite{Site-Reliability-Engineering-How-Google-Runs-Production-Systems}. However, manual system configuration, routine maintenance to ensure uptime, and extended troubleshooting processes have resulted in high operational costs and reduced customer satisfaction. Moreover, as monolithic systems transition to microservices architectures, systems face new challenges in managing complexity, ensuring fault recovery, and meeting scalability demands~\cite{Microservices-Yesterday-Today-and-Tomorrow}. To address these challenges, considerable efforts from both academia and industry have focused on developing the field of self-adaptive software systems, which aim to autonomously adjust system properties in response to changes in the operating environment.

To represent different operational requirements, the system properties are commonly divided into four principal properties of self-adaptiveness~\cite{the-vision-of-autonomic-computing}: (1) \textit{Self-configuration:} The ability of a system to automatically configure and reconfigure itself in response to changes without human intervention. Self-configuration is often triggered by events such as enabling new features, updating versions, or installing new components. (2) \textit{Self-healing:} The system’s capability to detect, diagnose, and recover from faults or failures. A self-healing system can either prevent failures or adjust its control flow to maintain the intended functionality. (3) \textit{Self-optimization:} The system's ability to adjust its performance and resource allocation to meet operational objectives. Common self-optimization techniques include scaling, dynamic load balancing, and resource scheduling. (4) \textit{Self-protection:} The system’s ability to proactively detect and mitigate security threats or attacks by addressing vulnerabilities or breaches. An example of self-protection is the ability to dynamically shift the system's attack surface to reduce exposure to threats.

Despite the promising aspects of self-adaptive software systems for managing complex software, teaching the subject in academia presents significant challenges~\cite{An-Educational-Course-on-Self-Adaptive-Systems-using-IBM-Technologies}. These challenges stem from the substantial conceptual knowledge required across this field and related disciplines, coupled with the hands-on learning necessary to achieve practical competence. The limited training available to students in this area has hindered the adoption of self-adaptive systems in industry. A survey of 355 industry practitioners revealed a shortage of skilled personnel with the necessary expertise to lead this topic~\cite{Self-Adaptation-in-Industry-A-Survey}. Notably, 64.4\% of industry practitioners expressed a need for greater support from the research community, particularly in self-optimization, metrics monitoring, and analysis when implementing self-adaptive systems.

In response to the education demands, we present our efforts in teaching a self-adaptive software systems course, detailing its structure and key components. The course is designed with several core elements, including hands-on learning, industry-relevant assignments, guest lectures from both industry and academic experts, and a final showcase featuring presentations, posters, and industry leaders as judges. This approach aims to equip students with theoretical knowledge, practical experience, and the necessary technical leadership skills.

Our course is designed with objectives along the following learning dimensions:

\begin{itemize}[leftmargin=*]
    \item \textbf{Knowledge Objectives}: This learning dimension is to provide students with a deep understanding of the theoretical and practical aspects of self-adaptive systems.
    The objectives are to: (a) grasp the conceptual model of self-adaptive software systems and their core self-* properties, (b) comprehend the fundamental techniques for designing self-adaptive systems with a closed feedback loop, and (c) get familiar with state-of-the-art solutions in self-adaptive systems and acquire understanding of current challenges.
    \item \textbf{Skills Objectives}: This dimension is focused on building the ability to practically apply the principles of self-adaptive systems. The objectives are to: (a) analyze and identify potential areas within a system where adaptive loops can be applied, (b) implement an appropriate adaptation approach to a given software system, and (c) understand the system behaviours and business objectives, and relate the self-adaptive techniques to achieve the desired behaviours.
    \item \textbf{Attitude Objectives}: This dimension aims to foster a collaborative, responsible, and forward-thinking mindset in dealing with complex, dynamic systems.
    The objectives are to: (a) develop the ability to work collaboratively within a team while taking responsibility for individual deliverables, (b) demonstrate the capability to clearly present and justify developed solutions, and (c) acknowledge the importance of self-adaptive systems in addressing runtime uncertainties and meeting quality of service (QoS) requirements.
\end{itemize}

To assess the effectiveness of our developed course, we use the Goal Question Metric (GQM) approach~\cite{goal-question-metric-approach} to systematically derive the following evaluation criteria, ensuring alignment between learning objectives and measurements:

\begin{itemize}[leftmargin=*]
    \item \textbf{EC1:} Do students demonstrate an improvement in their knowledge of self-adaptive systems compared to their understanding prior to taking the course?
    \item \textbf{EC2:} Are students able to apply at least one self-adaptive property to a practical software system upon completing the course?
    \item \textbf{EC3:} Can students actively relate self-adaptive software systems to their project development and identify areas where adaptive approaches can be applied?
    \item \textbf{EC4:} Do students find the novel component in our course -  showcase with industry judges, beneficial to the development of their technical and soft skills? 
\end{itemize}

The remainder of this paper is organized as follows. In Section 2 we review the related work. Section 3 outlines the course structure. In Section 4, we present the evaluation methods used to assess the effectiveness of the course, followed by the results and findings in Section 5. Section 6 discusses the key insights along with lessons learned. Section 7 discusses threats to validity. Section 8 provides the conclusions and future work.

\section{Background and Related Works}

Substantial efforts have been made by both academia and industry to promote the broader adoption of self-adaptive software systems. In this section, we review related works from industry applications, academic research, and educational initiatives, focusing on various aspects of advancing this field.

\subsection{Industry Developments}
Around two decades ago, IBM introduced the concept of autonomic computing in response to the increasing complexity of software environments~\cite{the-vision-of-autonomic-computing}. Autonomic computing refers to systems that manage themselves autonomously, guided by high-level objectives. Building on this vision, IBM launched the Autonomic Computing Toolkit~\cite{problem-determination-using-self-managing-autonomic-system} in 2004, which included embeddable components, tools, use case scenarios, and documentation. The toolkit aimed to apply adaptive mechanisms to software problem determination, installation, and integration solutions. Following IBM's success with self-adaptive software systems in commercial products, the open-source community began incorporating adaptive features into their offerings. For instance, Eclipse Equinox~\cite{eclipseEquinox}, a reference implementation of the Open Service Gateway Initiative (OSGi) framework, enabled modular components to be dynamically deployed, upgraded, and removed without downtime. Additionally, it provided resource monitoring and dynamic resource allocation to optimize performance.

In the early 2010s, the shift from monolithic architectures to microservices gained momentum. VMware’s 2011 release of the Cloud Foundry platform~\cite{cloudfoundry}, for example, introduced automatic runtime provisioning (using buildpacks) based on application language. Beyond self-configuration at deployment, Cloud Foundry extended these adaptive capabilities to runtime by enabling auto-reconfiguration via service binding. Java Spring applications, for instance, automatically adjust configurations based on Cloud Foundry environment variables.

As software architecture transitioned toward microservices, increasingly decoupled modules led to more advanced self-adaptive features. Platforms like Kubernetes~\cite{kubernetesDocumentation}, Google Borg~\cite{Large-Scale-Cluster-Management-at-Google-with-Borg}, and Red Hat OpenShift~\cite{openshiftDocymentation} began offering built-in functionalities for self-configuration through service discovery and self-optimization via horizontal and vertical scaling, allowing individual components to be resized dynamically. Self-healing features have also become widely adopted in industry software. For example, Microsoft Azure Security Center~\cite{microsoftDefender} provides adaptive authentication based on real-time risk assessments, while IBM QRadar~\cite{ibmQRadarSIEM} implements intelligent threat detection and response. These adaptive features have become integral to modern software applications, enabling them to self-manage and address specific operational challenges effectively.

\subsection{Research Activities}

While significant progress has been made in the development and implementation of self-adaptive software systems within the industry, the research community has made substantial contributions across several key areas, including theoretical foundations, architecture design, algorithm development, artifact creation, taxonomy, and roadmaps.

\textit{Theoretical Foundations:} Expanding on IBM’s initial vision for self-adaptive software systems, researchers have established foundational principles that emphasize the importance of capturing requirements during the development phase~\cite{Self-Adaptive-Software-Landscape-and-Research-Challenges}, which significantly influences the possible actions that could be taken during the operational phase. These principles also identify essential design considerations~\cite{Software-Engineering-for-Self-Adaptive-Systems-A-Second-Research-Roadmap}, such as observation, control, identification, and adaptation enactment, necessary for constructing effective self-adaptive systems.

\textit{Architecture Design:} A significant focus of research has been on architectural frameworks for self-adaptive systems. One common approach is the use of a closed-loop feedback control system~\cite{Self-Adaptive-Software-Landscape-and-Research-Challenges}, where the system continuously adjusts based on environmental and system feedback. A more sophisticated model is the MAPE-K loop~\cite{the-vision-of-autonomic-computing}, which includes \textit{Monitor} (monitoring the system), \textit{Analyze} (analyzing the collected data), \textit{Plan} (planning adaptive activities), and \textit{Execution} (executing the actions) components, along with an optional shared knowledge base (K). Layered architectures are also explored~\cite{self-managed-system-an-architecture-challenge}, where layers address different levels of adaptation granularity. For example, the top layer is the \textit{Goal Management} responsible for handling high-level requirements, the middle layer is \textit{Change Management}, which reacts to the change in states and is responsible for the execution of actions, and the lower layer \textit{Component Control} oversees sensor-actuator feedback mechanisms. The model-based approach, such as the Rainbow Framework~\cite{Rainbow-architecture-based-self-adaptation-with-reusable-infrastructure}, uses architectural models to monitor and inform adaptive decision-making processes.

\textit{Algorithm Development:} Alongside architectural advancements, numerous algorithms have been proposed to tackle specific challenges in self-adaptive systems, with a primary focus on managing uncertainty. Notable contributions include control-based approaches~\cite{SimCA-A-Control-theoretic-Approach-to-Handle-Uncertainty-in-Self-adaptive-Systems-with-Guarantees,Control-Strategies-for-Self-Adaptive-Software-Systems}, reinforcement learning~\cite{Adaptive-Action-Selection-in-Autonomic-Software-Using-Reinforcement-Learning,Online-reinforcement-learning-for-self-adaptive-information-systems}, multi-agent systems~\cite{self-adaptation-using-multiagent-systems,A-self-adaptive-multi-agent-system-approach-for-collaborative-mobile-learning}, and Bayesian networks~\cite{Taming-model-uncertainty-in-self-adaptive-systems-using-bayesian-model-averaging,Bayesian-artificial-intelligence-for-tackling-uncertainty-in-self-adaptive-systems-The-case-of-dynamic-decision-networks}, which have been pivotal in addressing issues like uncertainty in system adaptation.

\textit{Artifact Creation:} To support the exploration of self-adaptive system solutions, the research community has developed various artifacts as testbeds~\cite{Guidelines-for-Artifacts-to-Support-Industry-Relevant-Research-on-Self-Adaptation}. These artifacts pose specific software challenges that self-adaptive systems are expected to address. One well-known artifact is Znn.com~\cite{znncom}, which simulates the ``slashdot effect"~\cite{managing-flash-crows-on-the-internet} leading to a system overload due to abnormal traffic surges. Another example is AcmeAir~\cite{acmeAir}, a simulated airline booking system designed to mimic high scalability, cloud deployment, and multi-channel user interaction—representing a simplified version of real-world business applications with key business objectives.

\textit{Taxonomy and Roadmaps:} Research has also provided taxonomies and roadmaps to guide industry practitioners in implementing self-adaptive systems~\cite{Self-Adaptive-Software-Landscape-and-Research-Challenges,Software-Engineering-for-Self-Adaptive-Systems-A-Second-Research-Roadmap,Uncertainties-in-the-Modeling-of-Self-Adaptive-Systems-A-Taxonomy-and-an-Example-of-Availability-Evaluation,A-Survey-and-Taxonomy-of-Self-Aware-and-Self-Adaptive-Cloud-Autoscaling-Systems,On-decentralized-self-adaptation-lessons-from-the-trenches-and-challenges-for-the-futur,Generative-AI-for-Self-Adaptive-Systems-State-of-the-Art-and-Research-Roadmap}. These frameworks help capture system requirements during the development phase, address uncertainties, define adaptation mechanisms, and clarify the human role in such systems. Thus, offering a structured approach to the design and implementation of self-adaptive software. Through advancements in theory, design, and implementation, researchers have significantly contributed to establishing a robust foundation for building effective self-adaptive software systems.

\subsection{Education Courses }
With the growing body of research in self-adaptive software systems, several universities have integrated this topic into their curriculums. For instance, Litoiu at York University offers a course titled ``Engineering Adaptive Systems"~\cite{york-university-course-engineering-adaptive-systems} which emphasizes the theoretical foundations of self-adaptive software, including architecture design and runtime models. The course structure incorporates lectures, student-led paper presentations, and group discussions. Similarly, Weyns at Katholieke Universiteit Leuven teaches ``Engineering Self-Adaptive Software Systems"~\cite{linnaeus-niversity-engineering-self-adaptive-software-systems} covering fundamental principles and engineering aspects such as architecture-based adaptation, requirements engineering, and runtime modeling. The course features short lectures, discussion sessions, and peer feedback sessions to enhance student learning. At the University of Victoria, Muller offers a course on ``Self-Adaptive and Self-Managing Systems"~\cite{university-of-victoria-self-adaptive-and-self-managing-systems} focusing on engineering methodologies, architectures, algorithms, and techniques, with learning supported through research paper assignments and presentations.

While these courses provide a solid theoretical foundation, there is still a gap in the practical, hands-on engineering skills necessary to fully equip students to implement and lead self-adaptive software projects in industry settings.

\section{Course Structure}

Overall, the course structure can be divided into two main components--theory and practice. 

\begin{table*}[t]
\centering
\caption{12-Week Course Content and Schedule}
\label{tab:course-structure}

\setlength{\tabcolsep}{6pt} 
\renewcommand{\arraystretch}{1.25} 
\begin{tabular}{c p{4.5cm} p{8.5cm} p{3.15cm}}
\hline
\textbf{Week} & \textbf{Topic} & \textbf{Concepts Covered} & \textbf{Deliverables} \\
\hline

1 & Course Introduction & \\
\hline
2 & Principles of Self-Adaptation & 
Explain the basic principles of self-adaptation, describing the conceptual model of a self-adaptive system, and applying the conceptual model to a concrete self-adaptive application~\cite{A-Survey-and-Taxonomy-of-Self-Aware-and-Self-Adaptive-Cloud-Autoscaling-Systems, the-vision-of-autonomic-computing, Self-Adaptive-Software-Landscape-and-Research-Challenges} 
\\
& \textbf{Assignment Tutorial}: Local Setup &
Walk-through of the main technologies for assignments, which include Eclipse OpenJ9, Open Liberty, JMeter, and OpenShift \\
\hline
3 & Architecture of a Self-Adaptive System & 
Introduce a self-adaptive exemplar~\cite{deltaIoT} and cover essential maintenance tasks for automation as well as the primary functions of self-adaptation~\cite{the-vision-of-autonomic-computing}.
\\
& \textbf{Assignment Tutorial}: Cloud Setup & Walk-through of cloud deployment environment for the assignments, OpenShift on IBM Cloud and IBM Cloud Monitoring \\
\hline
4 & \textbf{Guest Lecture 1}: Session on OpenShift, OpenLiberty, IBM Semeru Runtimes, and Microservices & Provide an overview on container orchestration platform, Java runtimes, cloud application development and operation & Assignment 1 due\\
\hline
5 & Architecture-based Adaptation I & 
Cover the issues of architecture-based adaptation, degree of adaptability, and adaptation management~~\cite{Using-arch-models-for-runtime-adaptability, An-arch-based-aproach-to-sas} 
\\
\hline
6 & Architecture-based Adaptation II & Introduce the three layer architecture model, runtime architecture to realize adaptation, Rainbow Framework~\cite{Rainbow-architecture-based-self-adaptation-with-reusable-infrastructure}, reflection perspective on self-adaptation & Assignment 2 \& Quiz 1 due\\
\hline
7 & \textbf{Guest Lecture 2}: Adaptation in IBM Semeru Runtimes (Eclipse OpenJ9) & Deep dive into Java garbage collection and just-in-time compilation in Eclipse OpenJ9 & Course project proposal due\\ 
\hline
8 & Runtime Models I & Cover the topics of model-driven engineering, models and meta-models, reference model for self-adaptation, and runtime models~\cite{bencomo2014models} & \\
\hline
9 & Runtime Models II & 
Examine graph-based runtime adaptation~\cite{graf}, dynamic variability in complex adaptive systems~\cite{diva}, executable runtime megamodels~\cite{eurema} \\
\hline
10 & Requirements-driven Adaptation & Highlight the concepts of relaxing 
requirements, meta-requirements, functional 
requirements~\cite{Awareness-Requirements-for-Adaptive-Systems} & Assignment 3 due \\
\hline
11 & Control Theory-based Adaptation & Introduce the concepts behind feedback control loops, stability, accuracy, settling time, overshoot (SASO) properties, and overview of controller design~\cite{eng-sas-through-feedback-loops} as well as guarantees under uncertainty~\cite{Uncertainties-in-the-Modeling-of-Self-Adaptive-Systems-A-Taxonomy-and-an-Example-of-Availability-Evaluation} \\
\hline
12 & \textbf{Guest Lecture 3}: Reconciling High-accuracy, Cost-efficiency, and Low-latency of Inference Serving Systems & 
Present research efforts and challenges on autoscaling and model serving~\cite{ghafouri2024solution}   & Quiz 2 due \\
\hline
\end{tabular}
\end{table*}

\subsection{Theoretical Component}
In terms of theory, the course is designed based on weekly lectures to provide students with the fundamental concepts and approaches for engineering self-adaptive systems. The lectures are organized to cover a wide range of conceptual topics in a synchronous manner with practical components, as listed in Table~\ref{tab:course-structure}. The course content was developed based on key topics that have emerged within the research community over time~\cite{weyns2021selfadaptive}. The curriculum begins with the principles of self-adaptation, focusing on the core self-* properties. Next, the architecture of self-adaptive system is explored, with emphasis on closed-feedback loop and the MAPE-K model~\cite{the-vision-of-autonomic-computing}. Following this, the lectures delve into architecture-based and model-based approaches such as the Rainbow framework~\cite{Rainbow-architecture-based-self-adaptation-with-reusable-infrastructure}. Then, runtime models are studied as tools for enabling model-driven adaptation. Requirements-driven adaptation is also presented to illustrate methods of relating adaptation to high-level goals, such as quality-of-service objectives. Finally, control theory-based adaptation introduces mathematical models to optimize system behavior and manage uncertainty.

In addition to the weekly lectures, the theoretical component also features invited guest lectures, from both industry and academia, to provide different perspectives that help students relate theory to practice. The invited lecture from industry is geared towards providing additional insights into technologies related to the course assignments in the practical components of the course. The invited lecture from academia is focused on a deep dive into a state-of-the-art research approach, which aims to offer students inspiration relating to the implementation of their course project.

Lastly, the theoretical component of the course also includes two quizzes, which are delivered during the middle and end of the semester. The quiz format is structured based on open-ended case studies of well-known exemplars and established approaches from the research community. The objective of the quizzes is to evaluate student comprehension of theoretical concepts and foundational knowledge following the delivery of the course materials. In addition, these quizzes serve as a bridge between theory and practice. By engaging with well-known case studies in the quizzes, students develop a deeper understanding of key concepts, some of which could be later integrated into their course projects. This application of theoretical knowledge in a practical setting reinforces their learning and enhances their ability to analyze complex systems.

\subsection{Practical Component}

In terms of practice, the course is designed to offer students hands-on learning opportunities structured around three course assignments and one final course project, which can be completed individually or in groups.

\begin{itemize}[leftmargin=*]
\item {\textbf{Course Assignments}}: The goal of the course assignments is to offer a well-defined problem to help students put theory into action. At the same time, the course assignments also represent a great opportunity for students to gain hands-on experience with industry-relevant technologies. Thus, in collaboration with IBM, we purposefully designed the course assignments to be situated in the context of modern cloud computing. The primary technologies that make up the managed system and operating environment for the course assignments include the Eclipse OpenJ9 Java runtime~\cite{EclipseOpenJ9}, benchmark web applications implemented using Open Liberty framework~\cite{OpenLiberty}, and a container orchestration platform managed by OpenShift~\cite{openshiftDocymentation} on IBM Cloud. The desired outcome for the assignments is for students to apply learned concepts directly in the context of modern cloud computing technologies to create self-adaptive microservices, which is capable of meeting the demands of dynamic workloads (i.e., a custom auto-scaling solution). To help the students get started, a tutorial dedicated to providing a high-level overview of all relevant technologies is offered prior to the commencement of the assignments.

\quad To provide the students an opportunity for incremental learning, we divide the course assignments into three parts based on the MAPE-K reference model. The first assignment is focused on monitoring, the second assignment is focused on analysis, and the third assignment is focused on planning and executing. Each assignment is design to be a continuation of the prior assignment deliverables. The learning objective for the first assignment is for student to deploy the benchmark application in the operating environment and implement the monitoring component to collect metrics. The learning objective of the second assignment is for students to define adaptation goals and construct the analyzing component to identify metric trends that may indicate the need for adaptation. The learning objective for the third assignment is to realize a planning component to adaptation plans based on the analysis signals, implement the executing component to realize the plans, and close the adaptation loop.  

\quad After completing the assignments, students will have had an opportunity to build a self-adaptive system from the ground up. Moreover, they will have also gained various practical skills while working on the assignments in the context of industry-relevant technologies. 

\vspace{0.5em}

\item{\textbf{Course Project}}: Another key aspect of the practical component in this course is the course project. The course project represents an opportunity for students to openly explore self-adaptive solutions in domains of their own interest after the completion of the assignments. As a part of the development process, students are expected to submit an initial proposal highlighting the project goals, targeted self-* properties, as well as a preliminary overview of their self-adaptive system design. The final deliverable of the course project is a detailed report presenting their methodology and results obtained. 

\quad To conclude the course, an on-campus project showcase is also allotted at the end of the term. The showcase represents a unique feature in our course structure where students have the opportunity to showcase their projects in front of a panel of judges invited from both industry and academia. In return, the showcase offers the students an engaging environment to obtain valuable feedback on the outcomes of their course projects. The showcase can be divided into two main sessions that take place over the course of one day. During the first session, students take turns presenting their projects, in the format of 5-minute pitch talks, in front of their peers and the judge panel. During the second session, a dedicated poster area is set up for students to highlight in-depth details and results from their projects. Throughout these sessions, the invited judges are provided a set of rubrics for evaluating each project. For the pitch talks, the projects are rated based on communication, visual presentation, and technical content. For the poster session, the projects are graded based on organization and flow, scientific significance, visual appearance, and technical content. To provide extra motivation for the students, the top project based on this rubric is also awarded in the end. 

\quad In summary, the course project is designed such that students are encouraged to independently apply learned theory into practice, and gain additional hands-on experience when it comes to developing self-adaptive solutions.
\end{itemize}

\section{Course Evaluation}

As a part of on-going effort to deliver the best learning experience, two online surveys were implemented with the long-term goal of improving course quality. Overall, the surveys feature both Likert-scale questions as well as free-response questions to gather direct feedback from the students regarding their learning experiences. The overall rubric for the course evaluation can be broken down into the following categories:

\begin{itemize}[leftmargin=*]
    \item \textbf{Theoretical Knowledge}: The main objective of this evaluation dimension is to assess the dissemination of self-adaptive concepts to the class. The survey poses questions regarding the perceived helpfulness (from ``Not Helpful'', ``Somewhat Helpful'', to ``Very Helpful'') and difficulty (from ``Trivial'', ``Easy'', ``Moderate'', ``Challenging'', to ``Very Difficult'') of the quizzes as well as questions which ask the students to self-assess their knowledge of self-adaptive systems following completion of the course (from ``Basic'', ``Intermediate'', to ``Advanced'').
    \item \textbf{Hands-on Practice}: The focus of this dimension is on evaluating the design of course assignments. For example, the survey features questions focused on the relevancy of the assignment technologies and systems relative with respect to students' prior experiences (from ``Not Relevant'', ``Somewhat Relevant'', to ``Very Relevant'') as well as open-ended questions that give students the opportunity to share the learning obstacles encountered during each of the assignments.
    \item \textbf{Student-led Projects}: This area of the survey is focused on evaluating the impact of the course project on student learning. Questions under this category ask the students share their satisfaction level with respect to the outcome of their course project (from ``Unsatisfied'', ``Somewhat Satisfied'', to ``Very Satisfied'') and the amount of knowledge gained from completing the course project (from ``Did Not Improve'', ``Improved by a Little'', to ``Improved by a Lot'').
    \item \textbf{Showcase Learning}: This section of the survey is oriented towards collecting feedback on the unique showcase dedicated to showcasing student projects at the end of the course. Questions in this section of the survey ask students to provide open feedback on aspects of the showcase they enjoyed as well as aspects of the showcase that could use further improvement.
\end{itemize}

The surveys were delivered both during and after the completion of the course to gather comprehensive data points that help to evaluate the delivery of theoretical concepts and effectiveness of the practical components proposed in the course structure.

\section{Results and Findings}
In this section, we begin by presenting the demographic profile of our class, followed by an analysis of the survey results, which are organized into four key components of the course: theoretical knowledge, hands-on practice, student-led projects, and showcase learning. Each component aligns with a corresponding evaluation criterion.

\subsection{Demographics}
We evaluated the background of students across three dimensions: (1) work experience, (2) theoretical knowledge of self-adaptive software systems, and (3) familiarity with the technical components commonly used in such systems. 

In Table~\ref{table:student-work-experience} we provide a summary of students' work experience, classified into three categories: no work experience, internship or co-op experience, and full-time work experience. The vast majority of students have some level of industry exposure, through either internship from undergraduate studies or work prior to enrolling in the graduate program. Most of our students' work experiences are concentrated in application or platform development.

\begin{table*}[t]
\centering
\caption{Students' Industry Work Experiences}
\label{table:student-work-experience}
\begin{tabular}{l c p{13cm}}
\hline
\textbf{Experience Categories} & \textbf{\#} & \textbf{Representative Quotes} \\ \hline
No Industry Experience & 1 & ``No" \\ \hline
Internship or Coop & 6 & ``Infra and SDE Internships related to e-payment and automatic drive." \newline ``I focused on developing a high-voltage Battery Management System, involving hardware design, PCB design using EasyEDA, and implementation of a master-slave architecture." \\ \hline
Full-time Experience & 11 & ``Have one year full time working experience before, mainly work on recommendation systems and large scale distribution systems." \newline ``I am a software developer with 4+ years of work experience in full-stack application development, predominantly back-end development." \\ \hline
\end{tabular}
\end{table*}

We also included a Likert-scale question to evaluate students' level of familiarity with the subject matter, along with a binary question to determine whether they had previously encountered related topics in their academic curricula. These questions aimed to measure the students' foundational knowledge of self-adaptive software systems, providing a baseline for assessing the effectiveness of the course. As shown in Table~\ref{table:student-theoretical-knowledge-background}, the results indicate that only a small number of students had prior exposure to relevant concepts, with just two students reporting having limited experience in this field.

\begin{table}
\centering
\caption{Students' Theoretical Knowledge Background}
\label{table:student-theoretical-knowledge-background}
\begin{tabular}{p{6cm} c c}
\hline
\textbf{Background} & \textbf{\#} & \textbf{\%} \\ \hline
\textit{Have you previously learned about the topics and concepts related to self-adaptive systems in your academic curriculum?} & & \\
Yes & 5 & 27.8 \\ 
No & 13 & 72.2 \\ \hline
\textit{What was your level of familiarity with the concepts used during this course (e.g. MAPE-K, self-* properties, etc) prior to taking the course?} & & \\
Lots of experience working with these concepts & 0 & 0.0 \\ 
Limited experience working with these concepts & 2 & 11.1 \\ 
Heard of these concepts, but never worked with them & 5 & 27.8 \\  
Never heard of these concepts & 11 & 61.1 \\ \hline
\end{tabular}
\end{table}

\begin{figure}
    \centering
    \includegraphics[width=0.6\linewidth]{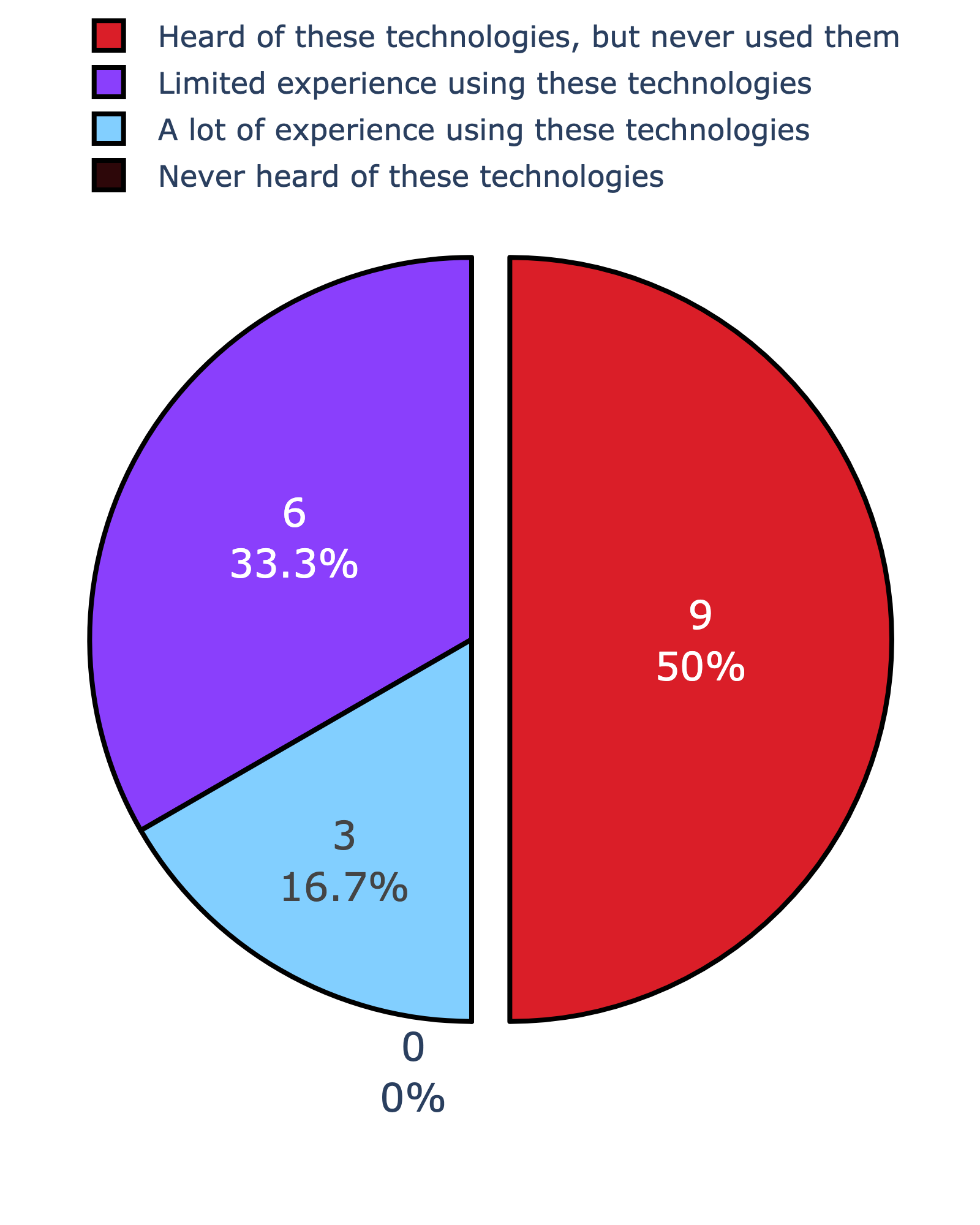}
    \caption{Students' familiarity with relevant technical components (i.e. IBM Cloud, OpenShift, Docker, etc)}
    \label{fig:background-students-familarity}
\end{figure}

Furthermore,  we evaluated students' technical experience with the components commonly used in self-adaptive software systems.
Students' responses in Figure~\ref{fig:background-students-familarity} show varying degrees of exposure to these technical components. Half of the students reported having direct, hands-on experience with these tools, while the other half shows awareness of these components.

\subsection{Theoretical Knowledge (EC1)}

To evaluate students' conceptual understanding of the theoretical foundations, we employed a pretest-posttest survey design~\cite{The-use-and-interpretation-of-quasi-experimental-studies} with an intermediate checkpoint to track knowledge progression. The intermediate survey was conducted after the completion of all assignments but before the project presentations in the showcase, and the final survey was taken after the course concluded. Due to the students' survey participation rate, we have three more responses in our final survey than in the intermediate survey result.

\begin{figure}
    \centering
    \includegraphics[width=0.9\linewidth]{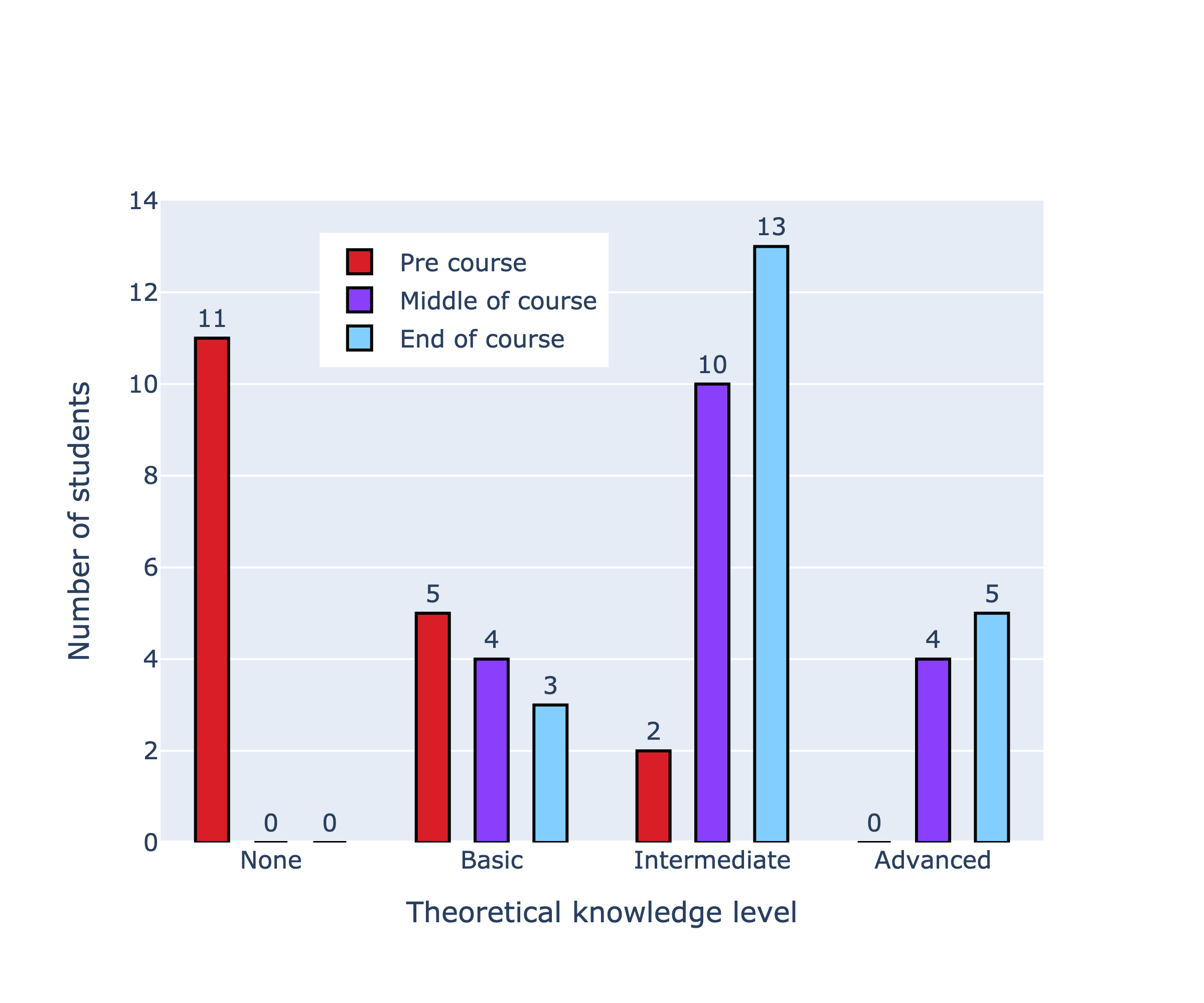}
    \caption{Students ranking their knowledge regarding self-adaptive systems at each progress point of the course}
    \label{fig:students-theoretical-knowledge-assessment}
\end{figure}

Figure~\ref{fig:students-theoretical-knowledge-assessment} aggregates the pre-intermediate-post survey results to rank students' knowledge of self-adaptive systems. A significant improvement in understanding was observed by the intermediate point of the course, where students had completed all theoretical material and successfully applied their knowledge in quizzes focused on identifying adaptive behaviours and models in a real-world software system.

\begin{figure}
    \centering
    \includegraphics[width=0.65\linewidth]{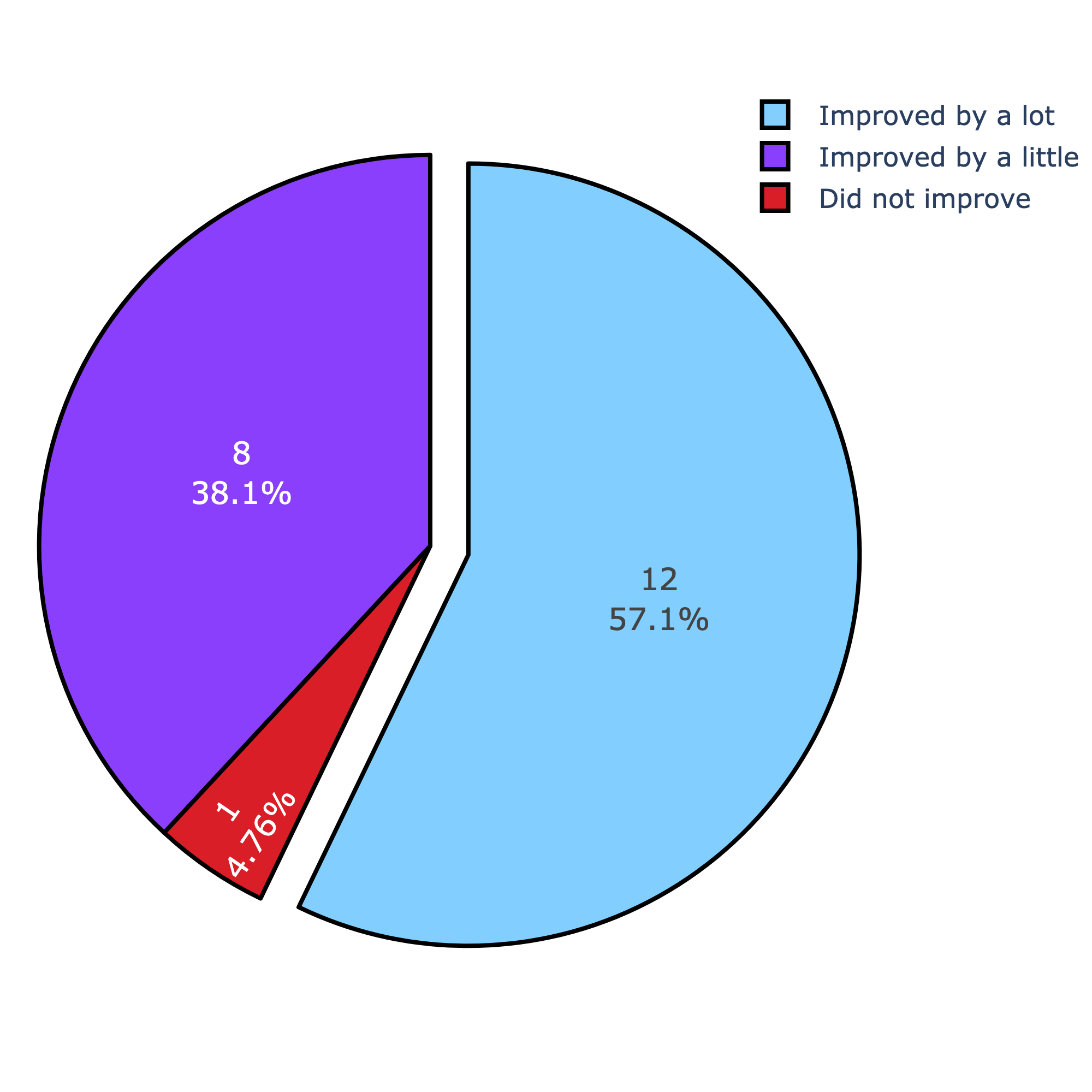}
    \caption{Survey question: \textit{How would you rank your knowledge regarding self-adaptive systems after completing the course project compared to before the course project?}}
    \label{fig:students-theoretical-knowledge-improvement}
\end{figure}

By the end of the course, following the completion of their project showcase, fewer students remained at the basic knowledge level despite more responses in our final survey, with a greater number demonstrating advanced understanding compared to the midpoint of the course. Additionally, when directly asked whether they felt their knowledge had improved, the majority of students reported substantial gains in understanding, as shown in Figure~\ref{fig:students-theoretical-knowledge-improvement}, indicating that the showcase contributed significantly to their knowledge growth. 

\begin{tcolorbox}[colback=gray!20, colframe=gray!80, title=Main Findings for EC1]
By the end of the theoretical learning and practice quiz, most students achieved an intermediate level of understanding. The project and showcase components considerably further enhanced their knowledge.
\end{tcolorbox}

\subsection{Hands-on Practice (EC2)}

To assess the effectiveness of the course in providing students with practical, real-world experience, we focused on two key dimensions: (1) the relevance of the software system used in the assignments to the industry systems students had previously encountered, and (2) whether the hands-on practice in the assignments enabled students to feel confident in applying any of the self-adaptive properties to a system in the future?

Top half of Table~\ref{table:student-practical-skills-improvement} presents the findings (from out of the 18 responses) regarding the relevance of the software system (AcmeAir) used in the assignments. The majority of students found the AcmeAir system to be highly relevant to industry systems they had previously worked with, suggesting strong alignment between the course content and real-world applications. Only one student reported that the system was not relevant to his/her prior experience.

Second half of Table~\ref{table:student-practical-skills-improvement} shows the number of students (out of 21 responses) who felt confident in applying each of the self-adaptive properties after completing the course assignments. Among the properties, self-configuration and self-optimization are the two top-ranked properties students feel confident in applying, with self-configuration having the most number of confidences. 

\begin{table}
\centering
\caption{Students' Hands-on Skills Improvement}
\label{table:student-practical-skills-improvement}
\begin{tabular}{p{6cm} c c}
\hline
\textbf{Survey Assessment} & \textbf{\#} & \textbf{\%} \\ \hline
\textit{To what degree do you think that the target system (AcmeAir) used in the course is relevant to the real-world systems you have experienced with in your previous work, internship or co-op experience?} & & \\
Very relevant & 11 & 61.1 \\ 
Somewhat relevant & 6 & 33.3 \\ 
Not relevant at all & 1 & 5.6 \\ \hline
\textit{Based on what you have learned through this course, which self-property (if any) you feel confident to apply in your future work?} & & \\
Self-configuration & 17 & 81.0 \\ 
Self-optimization & 15 & 71.4 \\ 
Self-healing & 7 & 33.3 \\  
Self-protection & 3 & 14.3 \\ \hline
\end{tabular}
\end{table}

These results indicate that the hands-on assignments successfully enhanced students' practical skills in applying key self-adaptive properties to industry-relevant systems. 

\begin{tcolorbox}[colback=gray!20, colframe=gray!80, title=Main Findings for EC2]
All students reported feeling capable of applying at least one self-adaptive property. The properties of self-configuration and self-optimization were the ones most students felt confident implementing, which is consistent with their prevalence in industry applications.
\end{tcolorbox}

\subsection{Students-led Projects (EC3)}

A key objective of our project component is to encourage students to independently think about and apply self-adaptive software system techniques with limited guidance. After completing guided assignments, where we provided a predefined system and specific areas for practice, students transitioned to a self-directed project. In this project, they were required to independently identify and implement appropriate techniques.

To evaluate students' ability to lead such implementations, we collected responses through open-ended survey questions. The results revealed a wide range of novel areas where students identified applications for self-adaptive software systems. Table~\ref{table:student-led-project} categorizes these responses and provides illustrative examples. The findings demonstrate that students developed a clear understanding of both the theoretical concepts and their projects, effectively applying self-adaptive techniques to meet key software requirements.

\begin{table*}[t]
\centering
\caption{Students Relating Self-adaptive Software System to Domain Knowledge}
\label{table:student-led-project}
\begin{tabular}{l c p{13cm}}
\hline
\textbf{Application Areas} & \textbf{\#} & \textbf{Representative Quotes} \\ \hline
System Behaviour/Functionality & 3 & ``Integrating self-adaptation into one of my older project named EmotionNet would significantly enhance its functionality by making the system more responsive and personalized to individual user needs. By tailoring content such as jokes, quotes, and YouTube links based on each user's emotional responses, EmotionNet could provide more effective emotional support. Additionally, optimizing the system to dynamically adjust its computational methods ensures smooth operation on the Raspberry Pi, even under varying loads. Overall, these self-adaptive capabilities would make EmotionNet not only more engaging but also more efficient in helping users manage their emotions." \newline ``Yes, one of my projects in a company involved writing a backend application which reserved a piece of data in two separate services. The problem was that my application did so in an imperative manner due to which there were situations where the data sometimes ended up reserving in one services while it got failed to reserve in the next service and too in a non-recoverable/rollback-able manner.
This led to strong inatomic consistencies in our ecosystem. One would expect atomicity in the form of either data getting completely reserved in both the services or getting the data rolled back to being in no services.
But something like that would have taken a lot of re-engineering. 
After studying this subject of self-adaptive systems, I believe I could have taken inspiration from the self-healing nature of self-adaptive systems to have a very simple isolated CronJob built which would run periodically and in every iteration, it would notice the data across both the services and reconcile/synchronize any drifts if found. This would have led to a self-healing eventually consistent ecosystem." \\ \hline
Resource Scheduling/Allocation & 4 & ``the load balancing problem that I used to simulate from a control course can be much more realistic and sophisticated with the knowledge of self-adaptation." \newline ``Yes. I can autoscale websites and optimize cloud resources usage." \\ \hline
Incident Handling/Maintenance & 3 & ``I personally work on payment systems every day we do need to constantly apply self-adaptive systems concepts to our applications to improve system reliability and uptime." \newline ``I would be using self-healing in my previous projects where we had periodic breakdown of the legacy backend environment would have saved me from a lot of sleepless nigths." \\ \hline
Other Application Areas & 7 & ``I had worked on a network security product in my past experience, we could apply the concepts in how we react to the events during network monitoring by the more structured approaches discussed in ESASS." \newline ``Most of the machine learning projects could benefit from the monitoring and self-adaptive feedback loop and in industry, MLOps is already being performed as a standard so yeah, it is very useful." \\ \hline
Unable to Identify an Area & 4 & ``Not quite, but it's majorly because my career path is more focus on another field." \newline ``N/A because of my limited work" \\ \hline
\end{tabular}
\end{table*}

\begin{tcolorbox}[colback=gray!20, colframe=gray!80, title=Main Findings for EC3]
Students demonstrated a solid understanding of relating self-adaptive software systems to their existing domain knowledge and identified novel areas of applications. 
\end{tcolorbox}

\subsection{Showcase Learning (EC4)}

The showcase with student-led project presentations and the involvement of industry practitioners as judges is a central components of our course structure. To bridge our students with industry and prepare them to lead the topic in their upcoming careers,  we invited five industry judges to participate in the showcase. Students first presented their project solutions to their peers and the judges, followed by interactive discussions during a poster session where the judges provided feedback.

\begin{figure}[b]
    \centering
    \includegraphics[width=0.65\linewidth]{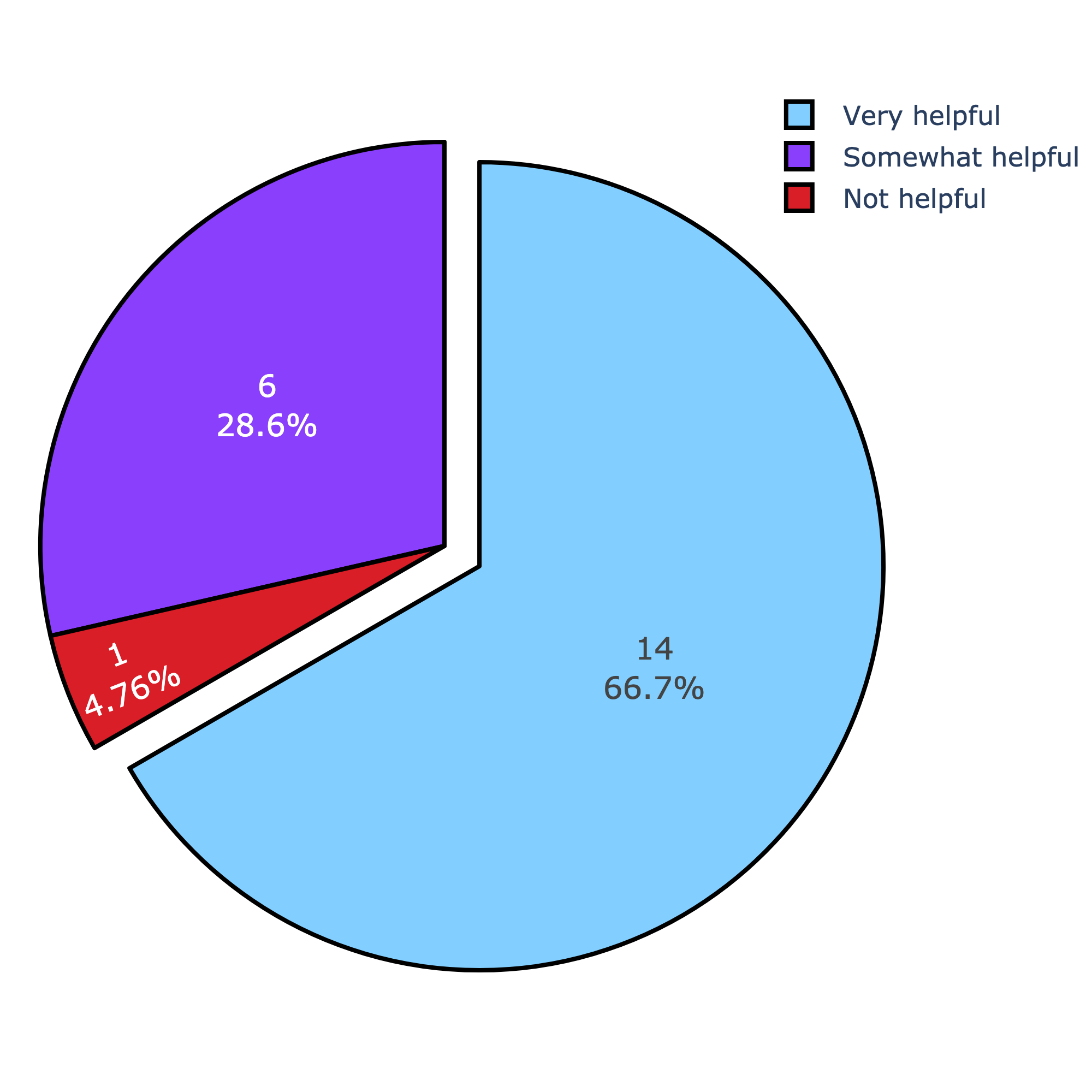}
    \caption{Survey Question: \textit{How helpful/beneficial did you find the course project + project showcase with respect to your learning outcome for this course?}}
    \label{fig:workshop-benefit-assessment}
\end{figure}

The primary goal of the showcase was to equip students with the necessary leadership skills in presenting technical solutions, allowing them to demonstrate knowledge in the field while learning from each other. We assessed the showcase’s effectiveness both quantitatively and qualitatively by addressing two key questions: (1) Did students find the project and the project showcase beneficial to their learning outcomes? (2) Which aspects of the showcase did students enjoy the most?

Our quantitative results are summarized in Figure~\ref{fig:workshop-benefit-assessment}, and the qualitative responses are presented in Table~\ref{table:students-feedback-on-workshop}. The majority of students found the showcase highly beneficial to their learning experience. From our analysis of student responses, we identified four key areas that students found most enjoyable: (1) presenting their ideas and solutions, (2) engaging with industry judges from IBM, (3) participating in the poster session where they interacted with peers, judges, and instructors, and (4) learning from their peers.

The opportunity to present their work gave students valuable practice in articulating technical solutions, an essential skill in both academics and industry. Interaction with industry judges provided students with real-world insights and practical feedback, bridging the gap between academic learning and industry application. The poster session provides a more relaxed environment for students to hold thorough discussions. Finally, the peer learning aspect helps to facilitate exchanges of ideas, gaining inspiration from others. 

\begin{table}
\centering
\caption{Students' feedback on the showcase}
\label{table:students-feedback-on-workshop}
\begin{tabular}{p{1.5cm} p{6cm}}
\hline
\textbf{Most perceived areas} & \textbf{Representative Quotes} \\ \hline
Presentation to class & ``Developing and deploying a practical solution, followed by a presentation to industry and academic leaders, was a pivotal experience. Collaborating with a teammate allowed us to effectively evaluate each other's strengths and weaknesses, facilitating the strategic allocation of tasks based on individual expertise. This dynamic mirrors the intricate process of real-world project management. The exposure gained from the showcase was invaluable, fostering healthy competition and providing a fitting culmination to our collective endeavors." \newline ``The technical-conference-like demos and a public stage to deliver our presentation. Moreover, this gave us the opportunity to have a very strong project in our portfolio." \\ \hline
Industry Interaction & ``1. Hands-on experience with tools that are used in industry. 2. Interacting and networking with senior industry professionals." \newline ``Pitching in front of the working professionals (IBM invitees and professors) and networking." \\ \hline
Learning from Peers & ``chance to see what peers did and learned" \newline ``I particularly enjoyed learning about other projects presented during the project showcase." \\ \hline
Poster Session & ``The poster session" \newline ``The poster" \\ \hline
\end{tabular}
\end{table}

\begin{tcolorbox}[colback=gray!20, colframe=gray!80, title=Main Findings for EC4]
The showcase was widely appreciated by the students. Its key components—presentations, industry interaction, poster sessions, and peer learning—together facilitate a unique 1-day learning experience. 
\end{tcolorbox}

\section{Lessons Learned}
Based on the findings presented in Section V. Results and Findings, and our experiences in teaching the course, this section offers insights for designing educational programs on self-adaptive software systems, along with broader recommendations for teaching industry-relevant yet challenging courses. While our experience is rooted in this specific topic, the lessons learned can be applied to other emerging areas, such as edge computing and machine learning in industry settings.

\subsection{Teaching Students with Diverse Backgrounds}
One of the key challenges in designing a course is aligning the teaching methods to accommodate a diverse student body with varying backgrounds and learning motivations~\cite{understanding-student-differences}. This is especially critical in our course, which requires substantial effort and technical competence from the students.

\textit{Lessons Learned:} Addressing student diversity was a primary consideration in course design. As our courses prepare the students with the skills to lead the relevant topics in industry settings, keeping all students motivated (including those who are interested in the topic but have a different career path) is a primary concern in developing the course~\cite{Software_Engineering_Education_Challenges_and_Perspectives}. We take the approach of Scaffolded Learning~\cite{scaffoldinglearning}, where we began by teaching foundational concepts to bring all students to a baseline level of understanding. For more experienced and interested students, supplementary materials were provided to deepen their knowledge. This approach ensured that all students could progress through the material, while those pursuing careers outside the subject area still gained valuable skills (e.g., soft skills like presenting and pitching). Survey results show promising results of this course design decision. For students with weaker background or planning on a different career path, they still found the course improved their knowledge significantly. 

\subsection{Challenges in Involving Substantial Industry Technologies}

With the rapid evolution of software technologies, researchers have identified several gaps between university education and the needs of the industry. One notable gap is in the use of technical tools~\cite{gapbetweenhighereducationandindustry}—most existing curricula focus on databases, object-oriented programming, while container technologies are rarely covered. Exposing students to the latest widely adopted software trends is essential for preparing competent software engineers~\cite{Perspectives_on_the_Gap_Between_the_Software_Industry_and_the_Software_Engineering_Education}. In our course, we explored the case of incorporating the latest industry technologies, including the OpenShift and IBM Cloud for all students, with IBM Watson Machine Learning available to those interested.

\textit{Lessons Learned:} We share two key insights from our experience integrating the latest tools into the course: \textit{1. Challenges for Students:} Students often struggle to adapt to new technologies and achieve the learning outcomes~\cite{Scalable_Teaching_of_Software_Engineering_Theory_and_Practice_An_Experience_Report}. To address this, we structured the course with progressively increasing difficulty. It began with an assignment introducing microservices and IBM Cloud services, followed by three guest lectures from industry and academia to provide deeper insights. For instance, before students tackled the analysis component in Assignment 2, a guest speaker from IBM covered the technical details of OpenShift, OpenLiberty, and Semeru Runtimes. Additionally, comprehensive technical documents and online tutorials from the cloud provider significantly helped students start. \textit{2. Challenges for Teaching Assistants (TAs):} Managing the infrastructure and providing timely support for students posed significant challenges for TAs. Efforts were required to set up the infrastructure and additional cloud services, as well as to respond promptly to student issues. Cloud services proved essential in streamlining infrastructure management. In our setup, TAs provided a single Kubernetes cluster to the students and isolated each team project within namespaces to prevent interference. This approach eliminated the need for local troubleshooting and allowed TAs to deliver on-demand assistance efficiently. Furthermore, we observed periodic fluctuations, with students working more intensively as deadlines approached, leading to a surge in infrastructure usage toward the end of each assignment and project, while usage remained noticeably lower at the start. The elasticity of cloud computing allowed TAs to effectively manage resource demand and control costs. Overall, the use of cloud services, combined with their scalability, ease of management, and extensive documentation, greatly facilitated the course's success.

\subsection{Balance Between Theoretical Learning and Practice}

The benefits of combining hands-on practical components with theoretical education have long been recognized by researchers~\cite{enhancing-learning-by-integrating-theory-and-practice}. However, within the limited timeframe of a semester, achieving solid outcomes in both presents a significant challenge~\cite{Experience_of_Teaching_Data_Visualization_using_Project-based_Learning}.

\textit{Lessons Learned:} To address this, we divided our course into two parts. The first half focused on theoretical learning through lectures, complemented by assignments designed to develop practical skills. In the second half, students engaged in a student-led project, allowing them to explore areas of interest and apply theoretical concepts to a specific domain. Our assessments show that this approach effectively helps both theoretical and practical skill development. However, a common feedback from students was to introduce the project earlier in the course, allowing more time for idea development. We plan to incorporate this suggestion in the next iteration by starting the project brainstorming phase earlier between Assignment 2 and Assignment 3. Additionally, for students with less software experience, we plan to offer dedicated office hours to help them explore areas in which self-adaptive theories and concepts can be applied to their project ideas.

\section{Threats to Validity}
In this section, we discuss the potential threats to the validity of our study and the actions we took for mitigation. 

\subsection{Construct Validity}

In this study, we employed two surveys as proxies to measure students' understanding of the subject and their perceptions of the course's effectiveness. However, self-reported data is susceptible to response bias~\cite{response-bias}, as students may be inclined to select answers that align with social expectations, particularly in Likert-scale questions. To mitigate this bias, we implemented anonymous online surveys and distributed the final survey only after course grades were released, ensuring that students did not perceive the survey as influencing their academic evaluation. Additionally, to complement the Likert-scale questions, we included qualitative questions requiring detailed responses to provide a comprehensive assessment.

\subsection{Internal Validity}

We recognize several confounding factors that may threaten the internal validity. Students may have acquired knowledge from external resources outside the course content or through interactions with peers during the group project. Additionally, our results are subject to selection bias~\cite{selection-bias}, as not all students participated in the surveys. Specifically, we received 18 responses for the first survey and 21 for the second, out of a total of 23 students. Those who performed better in the course might be more likely to participate in the survey.

\subsection{External Validity}

Two threats to external validity arise from the limited number of students in the reported course offering and the specific industry technologies used. The findings are based on a single course offering with 23 students in one country, which may not be representative of a broader population. As a pioneer course in teaching self-adaptive systems, we provide student demographics to support the potential replication of this study. Additionally, the course utilized industry technologies such as OpenShift and IBM Cloud for its infrastructure. The findings may not generalize to other technologies due to differences in documentation maturity and usability challenges.

\section{Conclusions and Future Work}

Teaching a course that balances theoretical understanding and practical skills is often a challenging task. In this work, we shared our experience teaching self-adaptive software systems, covering key components and historical advancements in the field, alongside hands-on assignments to enhance practical learning. The final project and showcase further deepened students' knowledge, as evidenced by survey results. To close the gap between academic education and industry practices, we incorporated guest lectures and industry judges. Feedback from survey results also indicates that students successfully achieved the learning objectives. In future iterations, we plan to integrate student suggestions for improvement. Our course structure can also be adapted for education in other emerging software fields.

\section*{Acknowledgment}

The authors of this paper would like to express gratitude for the invaluable support from IBM in the development of this course, including IBM members Vio Onut, Kishor Patil, Vij Singh, and Marcellus Mindel. We also especially thank Andrew Craik for his contributions towards the design of this course. In addition, we would like to extend our appreciation to all of our students for their active participation in the class learning and their invaluable feedback. We also appreciate the insightful comments on this paper by the CSEE\&T reviewers.

\end{document}